\documentclass[twocolumn]{revtex4-1}

\usepackage[dvips]{color}
\usepackage{graphicx}
\usepackage{multirow}
\usepackage[]{amsmath}

\begin{document}

\author{H. Rodrigues} \email{E-mail address: harg@cefet-rj.br}
\affiliation{Departamento de F\'isica, Centro Federal de Educa\c{c}\~ao Tecnol\'ogica Celso Suckow da Fonseca \\
 Av Maracan\~a 229, 20271-110, Rio de Janeiro, RJ, Brazil}

\author{A. M. Endler} \author{S. B. Duarte} \email{E-mail address: sbd@cbpf.br} 
\affiliation{Centro Brasileiro de Pesquisas F\'isicas \\
 Rua Dr. Xavier Sigaud 150, 22290-180, Rio de Janeiro, RJ, Brazil}

\author{M. Chiapparini} 
\affiliation{Instituto de F\'isica, Universidade do Estado do Rio de Janeiro \\
 Rua S\~ao Francisco Xavier 524, 20550-900, Rio de Janeiro, RJ, Brazil}

\title{Gravitational wave generation from dynamical shape transition of protoneutron stars}

\begin{abstract} 

We describe the dynamical behavior of newborn pulsars modeled as homogeneous rotating spheroids. The dynamical evolution is triggered by the escape of trapped neutrinos, provided the initial equilibrium configuration. It is shown that for a given set of values of the initial angular momentum a shape transition to a triaxial ellipsoid configuration occurs. Gravitational waves are then generated by the breaking of the axial symmetry, and some aspects of their observation are discussed. We found a narrow window for the initial values of the angular frequency and the  eccentricity able to  enable a dynamical shape transition, with the Kepler frequency of the rotating fluid determining the upper bound of the initial angular frequency and eccentricity. The loss of energy and angular momentum carried away by the gravitational wave is treated consistently with the solution of the equations of motion, which govern the dynamical evolution of the system. The transition in the shape of the core of the pulsar is weakly dependent on the time scale of the neutrino escape. 

\begin{description}
\item[keywords] gravitational waves; rotating neutron stars; compressible ellipsoids
\item[PACS] 95.30.Sf;04.30.-w;97.60.Jd
\end{description}

\end{abstract}


\maketitle

\section{Introduction}\label{sec:intro}

When a massive star reaches the end of its long evolutionary process, a dense core  composed of nuclei of the iron group is formed. Photo-dissociation of nuclei and electron capture lower the hydrostatic pressure, and thus the core collapses \citep{Fridolin}. 

During the gravitational collapse, most of the gravitational binding energy ($\sim 10^{53}\ {\rm erg}$) is used to produce a huge amount of neutrinos ($\sim 10^{57}$). The rest of the gravitational energy is converted into kinetic energy of the ejected stellar material and the generation of gravitational waves and electromagnetic radiation. Depending on the mass of the core star, the gravitational collapse can lead to the formation of a neutron star in the aftermath of the supernova explosion \citep{Glendenning}. The core of nascent neutron stars may be very dense, rotating with a short regular period ranging from seconds to a few milliseconds. Rotation may induce  non-spherical deformations, and according to Theory of General Relativity if the mass distribution has a quadrupole component, part of the energy of the system is supposed to be converted into gravitational radiation \citep{Ohanian}. As important sources of gravitational waves, the study of neutron stars seems to be relevant in designing modern detectors of gravitational waves \citep{Riles}. 

A rigorous analysis of the protoneutron star formation is a complex task, requiring the use of elaborate numerical code for relativistic hydrodynamic as well as the input of a huge amount of physical data not well established yet. In addition, a compatible gravitational wave generation mechanism with the lost of angular momentum and energy carried away by the gravitational emission should be considered. So, the equations governing the lost of those quantities have to be coupled to the dynamical calculation. 

Many different proposals have been already presented in the literature to clarify the question of gravitational wave generation in a newborn pulsar. Different normal modes of oscillations in the fluid medium has been explored and discussed, claiming their excitation by glitches produced in the pulsars structure \citep{Abadie}. The possibility of the r-modes excitation have recently deserved more attention \citep{Alford} due to the compatibility of the range of frequency and the order of magnitude of the amplitude of the generated gravitational wave with the observability of new generation detectors projects. Such oscillation mode was introduced for the first time more than a decade ago \citep {Andersson,Friedman}. Recently some authors pointed out the existence of a critical rotational frequency of the inner fluid in the contact with a solid crust to have formed the r-mode oscillations \citep{Alford2}. All modes of fluid oscillation are supposed to be attenuated in time due to the viscous rotating flow. 

In the absence of a satisfactory model to address the problem of generation of gravitational waves in the context of the formation of the neutron stars, it may be promising work with models connecting general aspects of the emitted gravitational waves,  and the energy of the excited modes in the system. Faced this intricate scenery, simplified dynamical models which are able to capture basic aspects of the problem can be still an useful tool to explore alternative explanations of the generation mechanism. 

In this article we model rotating proton neutron star as homogeneous configurations having ellipsoidal shape. We derive an effective Lagrangian of the system including terms of the kinetic energy, rotational energy, the gravitational potential energy, and the internal energy, defined as functions of the semi-axes of the ellipsoid and their time derivatives. The gravitational wave radiated by the protoneutron star is treated in the framework of quadrupole approximation of Einstein equations, determining  the intensity of the radiated gravitational waves. The equations of motion are obtained and solved numerically for a given initial equilibrium of spheroidal configuration. Considering the radiation of gravitational waves we have additional equations coupled to the semi-axes equations of motion, one of them describing the rate of the radiated energy and the other one describing the rate of the angular momentum loss. 

The initial configurations are constructed in order to represent the core of a rotating protoneutron star as a equilibrated homogeneous spheroid configuration. We consider the neutron star matter composed of neutrons, protons, hyperons, delta resonances and relativistic electrons and muons described in the framework of a relativistic mean field theory. 
   

The work is organized as follows. In Section \ref{sec:model}, we derive the equations of motion describing the dynamics of compressible triaxial ellipsoids. In Section \ref{sec:gwaves} we briefly establish the equations governing the emission of gravitational waves in the framework of the weak field limit. In Section \ref{sec:neutrinos} we present the  simplified description for the neutrino escape, aiming to discuss the effect of the time scale of neutrinos on the dynamical condition  obtained for the system to become gravitational waves emitter. In Section \ref{sec:result}, we present and discuss the obtained results. Finally, in Section \ref{sec:conclu} we summarize the main conclusions.

\section{Description of the model}\label{sec:model}

The dynamics of a uniformly rotating neutron star is approximated by a compressible homogeneous triaxial ellipsoid.  We construct an effective Lagrangian
\begin{equation}
{\cal L} = K  - W - U_{rot} - U_{int},
\end{equation}
with $K$ being the translational kinetic energy, $W$ the gravitational potential energy, $U_{int}$ the internal energy, and $U_{rot}$ the rotational kinetic energy (the {\it centrifugal potential}).  All quantities are written in terms of the semi-axes of the ellipsoid $a_1$, $a_2$, and $a_3$, and their respective time derivatives. 

The gravitational potential energy of the triaxial ellipsoid is given by \citep{Chandra}
\begin{equation}
W=-\frac{3}{10}GM^{2}\frac{A}{a_{1}a_{2}a_{3}},  \label{19}
\end{equation}%
where $A$ is defined by 
\begin{equation}
A=\sum_{i=1}^{3}A_{i}a_{i}^{2}, \, \, \, \, i=1,2,3 \label{4}
\end{equation}%
where 
\begin{equation}
A_{i}=a_{1}a_{2}a_{3}\int_{0}^{\infty }\frac{d\zeta }{\Delta \left(
a_{i}^{2}+\zeta \right) },  \label{3}
\end{equation}%
with
\begin{equation}
\Delta ^{2}=(a_{1}^{2}+\zeta )(a_{2}^{2}+\zeta )(a_{3}^{2}+\zeta ).
\label{5}
\end{equation}

The coefficients $A_{ì}$ regarding the six possible triaxial ellipsoidal
configurations are given by 
\begin{equation}
A_{i}=\frac{2a_{j}a_{k}}{a_{i}^{2}k^{2}u^{3}}\left[
F(u,k)-E(u,k)\right] , \label{8}
\end{equation}
\begin{eqnarray}
A_{j}&=&\frac{2a_{j}a_{k}}{a_{i}^{2}k^{2}(1-k^{2})u^{3}} \nonumber \\
&& \times \left[
E(u,k)-(1-k^{2})F(u,k)-\frac{a_{k}}{a_{j}}k^{2}u\right] ,
\label{9}
\end{eqnarray}%
and%
\begin{equation}
A_{k}=\frac{2a_{j}a_{k}}{a_{i}^{2}(1-k^{2})u^{3}}\left[ \frac{a_{j}}{a_{k}}%
u-E(u,k)\right] ,  \label{10}
\end{equation}%
where all permutations are constrained by the condition
$a_{i}>a_{j}>a_{k}$, with the definitions
\begin{equation}
\alpha =\arccos \frac{a_{k}}{a_{i}},\ \ \ \ u=\sin \alpha ,
\end{equation}
and
\begin{equation}
k^{2}=\frac{a_{i}^{2}-a_{j}^{2}}{a_{i}^{2}-a_{k}^{2}}=\frac{%
1-a_{j}^{2}/a_{i}^{2}}{u^{2}}.  \label{11}
\end{equation}

The functions%
\begin{equation}
F(u,k)=\int_{0}^{u}\frac{dt}{(1-t^{2})^{\frac{1}{2}}(1-k^{2}t^{2})^{\frac{1}{%
2}}},  \label{12}
\end{equation}%
and%
\begin{equation}
E(u,k)=\int_{0}^{u}\frac{(1-k^{2}t^{2})^{\frac{1}{2}}}{(1-t^{2})^{\frac{1}{2}%
}}dt,  \label{13}
\end{equation}%
are the incomplete elliptic integrals of the first kind and the second
kind, respectively.

The coefficients $A_i$ satisfy the relationship 
\begin{equation}
\sum_{i=1}^{3}A_{i}=2. 
\end{equation}

The fluid motion inside the ellipsoid is governed by the continuity equation
\begin{equation}
\nabla \cdot \vec{v}=-\frac{\dot{\rho}}{\rho }.  \label{20}
\end{equation}%
For an irrotational flux the velocity field in the rotating frame is given by
\begin{equation}
\vec{v}=\sum_{i=1}^{3}\frac{\dot{a}_{i}}{a_{i}}X_{i}\hat{e}_{i},
\label{21}
\end{equation}%
where each component is a linear function of the coordinates. Due to the linear dependence of the velocity field on the
coordinates, the ellipsoidal boundary of the system is preserved every time. 

The translational kinetic energy associated with the internal motion  
is computed by the integral%
\begin{equation}
K=\int_{V}\frac{1}{2}\rho \left\vert \vec{v}\right\vert ^{2}dV.
\label{22}
\end{equation}

Substituting equation (\ref{21}) and carrying out the integral in the right-hand
side of equation (\ref{22}), one obtains the simplified quadratic form
\begin{equation}
K=\frac{1}{10}M\left(
\dot{a}_{1}^{2}+\dot{a}_{2}^{2}+\dot{a}_{3}^{2}\right) .
\label{23}
\end{equation}

The rotational energy measured in the rest frame fixed at the center of the ellipsoid is given by 
\begin{equation}
U_{rot}=\frac{1}{2}I\Omega ^{2},  \label{24}
\end{equation}%
where $I$ is the moment of inertia of the ellipsoid relative to
the axis of rotation, which reads
\begin{equation}
I=\frac{1}{5}M(a_{1}^{2}+a_{2}^{2}).  \label{25}
\end{equation}

The angular velocity can be expressed in terms of the 
angular momentum $L$ of the ellipsoid, $ L=I \Omega$, and thus we may put the rotational energy
(\ref{24}) in the form
\begin{equation}
U_{rot}=\frac{5}{2}\frac{L^{2}}{M(a_{1}^{2}+a_{2}^{2})}.
\label{27}
\end{equation}

The equations of motion for the three semi-axes obtained from the Lagrangian of the system are thus given by
\begin{equation}
\ddot{a}_{1} = - \frac{3}{2}\frac{GM}{a_{2}a_{3}}A_{1}+\frac{25 L^{2}}{M^{2}}%
\frac{a_{1}}{(a_{1}^{2}+a_{2}^{2})^{2}}+\frac{20\pi }{3M}Pa_{2}a_{3}, \label{29}
\end{equation}
\begin{equation} 
\ddot{a}_{2} = - \frac{3}{2}\frac{GM}{a_{1}a_{3}}A_{2}+\frac{25 L^{2}}{M^{2}}%
\frac{a_{2}}{(a_{1}^{2}+a_{2}^{2})^{2}}+\frac{20\pi }{3M}Pa_{1}a_{3},
\label{30}
\end{equation}
and
\begin{equation}
\ddot{a}_{3}  = - \frac{3}{2}\frac{GM}{a_{1}a_{2}}A_{3}+\frac{20\pi }{3M}%
Pa_{1}a_{2},  \label{31}
\end{equation}
where $P$ is the fluid pressure.

From equations (\ref{29}) and (\ref{30}) we obtain the equation
\begin{equation}
a_{1} \ddot{a}_{1} - a_{2} \ddot{a}_{2}=  2\pi G\rho
(A_{2}a_{2}^{2}-A_{1}a_{1}^{2})-\Omega ^{2}(a_{2}^{2}-a_{1}^{2}).
\label{34}
\end{equation}

The hydrostatic equilibrium entails that $\ddot{a}_{1}=\ddot{a}_{2}=\ddot{a}_{3}=0$, which from (\ref{34}) leads to
\begin{equation}
\frac{\Omega ^{2}}{\pi G\rho }=2\frac{A_{1}a_{1}^{2}-A_{2}a_{2}^{2}}{%
a_{1}^{2}-a_{2}^{2}}.  \label{35}
\end{equation}%

Equation (\ref{35}) represents the Jacobi equation for triaxial ellipsoids in equilibrium \citep{Chandra}.
On the other hand, from equations (\ref{29}) and (\ref{30}) we find out
\begin{equation}
2\pi G\rho \left( A_{2}-A_{1}\right) a_{1}^{2}a_{2}^{2}=\frac{20\pi }{3M}%
Pa_{1}a_{2}\left( a_{1}^{2}-a_{2}^{2}\right) a_{3}.  \label{36}
\end{equation}%
But from equation (\ref{31}) we get
\begin{equation}
\frac{20\pi }{3M}Pa_{1}a_{2}=2\pi G\rho A_{3} a_{3},  \label{37}
\end{equation}
which gives
\begin{equation}
2\pi G\rho \left( A_{2}-A_{1}\right) a_{1}^{2}a_{2}^{2}=2\pi G\rho
a_{3}^{2}A_{3}\left( a_{1}^{2}-a_{2}^{2}\right) ,  \label{38}
\end{equation}%
and hence%
\begin{equation}
a_{1}^{2}a_{2}^{2}\frac{A_{2}-A_{1}}{a_{1}^{2}-a_{2}^{2}}=a_{3}^{2}A_{3}.
\label{39}
\end{equation}

The geometric relationship in equation (\ref{39}) may also be represented by 
\begin{equation}
A_{1}-\frac{a_{3}^{2}}{a_{1}^{2}}A_{3}=A_{2}-\frac{a_{3}^{2}}{a_{2}^{2}}
A_{3}.  \label{40}
\end{equation}

\section{The gravitational wave emission}\label{sec:gwaves}

The quadrupole moment $Q_{\alpha\beta}$ of a distribution of mass in the fixed inertial frame is defined by the symmetric tensor \citep{Ohanian,Riles}
\begin{equation}
Q_{\alpha \beta }\equiv \int \rho (\vec{x})\left( 3x_{\alpha
}x_{\beta }-\delta _{\alpha \beta }x_{\gamma }^{2}\right) d^{3}x,
\label{43}
\end{equation}%
where $\rho (\vec{x}) $ is the mass density at the point of coordinates $x_i$ relative to the fixed frame. 
 
In the framework of weak field approximation, the time rate of energy carried by the gravitational wave is given by \citep{Landau}
\begin{equation}
 \frac{dE}{dt} = -\frac{G}{45c^{5}} \left( \frac{\partial ^{3}Q_{\alpha \beta }}{
\partial t^{3}}\right)^{2},  \label{52}
\end{equation}
from which we can obtain the mean value of the gravitational wave luminosity 
\begin{equation}
L_{\rm GW} =-\left\langle \frac{dE}{dt} \right\rangle.  \label{52d}
\end{equation}

For a homogeneous triaxial ellipsoid rotating uniformly about the $x_{3}$-axis with the frequency $\Omega$, the components of the quadrupole moment $Q_{\alpha \beta }$ are explicitly given by
\begin{eqnarray}
Q_{11} &=&Q_{11}^{\prime }\cos ^{2}\Omega t+Q_{22}^{\prime }\sin^{2}\Omega
t,  \label{47} \\
Q_{22} &=&Q_{11}^{\prime }\sin^{2}\Omega t+Q_{22}^{\prime }\cos^{2}\Omega
t,  \label{48} \\
Q_{12} &=& Q_{21}=\frac{1}{2}\left( Q_{11}^{\prime }-Q_{22}^{\prime }\right) \sin 2\Omega t,  \label{49} \\
Q_{33} &=&Q_{33}^{\prime },  \label{50}
\end{eqnarray}%
and $Q_{13}=Q_{13}=Q_{23}=Q_{32}=0$, where the quadrupole moment tensor $Q_{ij}^{\prime}$ determined in the rotating frame
reads%
\begin{equation}
Q_{ij}^{\prime }=\left\{ 
\begin{array}{ll}
\frac{1}{5}M\left( 3a_{i}^{2}-a_{\alpha }a_{\alpha }\right) , & {\rm for } \, \, 
i=j \\ 
0, \, \,  {\rm otherwise.}%
\end{array}%
\right.   \label{51}
\end{equation}

The equations (\ref{47}) and (\ref{48}) can be put in the form  
\begin{eqnarray}
Q_{11} &=&\frac{1}{2} \left(Q_{11}^{\prime} + Q_{22}^{\prime} \right) + \frac{1}{2} \left(Q_{11}^{\prime} - Q_{22}^{\prime} \right) \cos 2 \Omega t,  \label{47b} \\
Q_{22} &=&\frac{1}{2} \left(Q_{11}^{\prime} + Q_{22}^{\prime} \right) - \frac{1}{2} \left(Q_{11}^{\prime} - Q_{22}^{\prime} \right) \cos 2 \Omega t.  \label{48b}
\end{eqnarray}



By using equations (\ref{51}) to (\ref{48b}) together with equation (\ref{52}), we get from equation (\ref{52d}) the explicit expression of the gravitational luminosity \citep{Peters,Chin,Chau,Beltrami}
\begin{equation}
L_{\rm GW} = \frac{32}{125}\frac{GM^{2}\Omega ^{6}}{c^{5}}\left(
a_{1}^{2}-a_{2}^{2}\right) ^{2}.  \label{61}
\end{equation}

The rate of angular momentum loss due to the gravitational radiation can also be determined \citep{Peters}:
\begin{equation}
\frac{dL_{i}}{dt}=-\frac{2G}{45c^{5}}\epsilon _{ijk}\frac{\partial ^{2}Q_{jl}%
}{\partial t^{2}}\frac{\partial^{3}Q_{kl}}{\partial t^{3}},  \label{62}
\end{equation}%
where $L_{i}$ is the $i$-th component of the angular momentum $\vec{L}$ and $%
\epsilon _{ijk}$ stands for the Levi-Civita symbol. If the rotation is about the axis
of symmetry, the $x_{3}$-axis component is the sole nonzero component (that is, $L_3 = L$), and in this case we get
\begin{equation}
\frac{dL}{dt}=-\frac{32}{125}\frac{GM^{2}\Omega ^{5}}{c^{5}}\left(
a_{1}^{2}-a_{2}^{2}\right) ^{2}.  \label{63}
\end{equation}

Alternatively, from equation (\ref{61}) we can derive the following equation 
for the angular velocity:
\begin{eqnarray}
\frac{d\Omega }{dt}&=&-\frac{1}{a_{1}^{2}+a_{2}^{2}} \nonumber \\
& \times & \left[ \Omega \left( a_{1}%
\dot{a}_{1}+a_{2}\dot{a}_{2}\right) +\frac{32}{25}\frac{GM\Omega ^{5}}{c^{5}}%
\left( a_{1}^{2}-a_{2}^{2}\right) ^{2}\right]  \label{velocidadeangular}
\end{eqnarray}
that must be solved along with the equations of motion (\ref{30}) to (\ref{31}). 
  
 For large distances between the emitting source and the observer, the waves can be taken as plane. So, we suppose that the gravitational energy is carried by a plane wave of amplitude $h_0$ and frequency $\omega$. The mean energy flux $\left\langle F \right\rangle$ and the latter quantities are then related by \citep{Landau}
\begin{equation}
\left\langle F \right\rangle= \frac{c^3}{32 \pi G} h_0^2 \omega^2 \label{flux1},	
\end{equation}
where the symbol $ \left\langle \cdot \right\rangle $ stands for the time average.

The gravitational luminosity and the mean flux observed at a point located at a distance $r$ from the emitting source are connected by
\begin{equation}
L_{\rm GW} = \int_S \left\langle F \right\rangle dS, 
\end{equation}
which gives 
\begin{equation}
\left\langle F \right\rangle = \frac{ L_{\rm GW}}{4 \pi r^2}. \label{flux2} 
\end{equation}
  
So, making use of equations (\ref{flux1}) and (\ref{flux2}) we obtain for the wave amplitude
\begin{equation}
h_{0}^{2}=\frac{8G}{\omega^{2}c^{3}}\frac{L_{\rm GW}}{r^{2}}. \label{amplitude1}  
\end{equation}

Equations (\ref{47b}) and (\ref{48b}) entail that the wave frequency is $\omega = 2 \Omega$, that is, it is twice the frequency of rotation of the emitting source \citep{Chin}. So, the wave amplitude can be estimated as 
\begin{equation}
h_{0}=4.5856\times 10^{-22}\left( \frac{G L_{\rm GW}}{c^{3} } \right)^{1/2}
\frac{1}{\Omega r^*} \ [{\rm cm}] ,
\end{equation}
where $r^*$ is the distance from the source and the Earth, in units of kpc.

\section{Neutrino escape parametrization}\label{sec:neutrinos} 

In the final instants of the gravitational collapse,  neutrinos are trapped within the  supernova core \citep{Rodrigues}. However, during the first instants of the protoneutron star formation, the trapped neutrinos begin to leave the core, carrying part of the system energy away. 

In order to model the escape of neutrinos, we assume that neutrinos leave the system at a constant rate, that is,
\begin{equation}
\frac{d N_{\nu}}{dt} = - \tau , \label{dNdt}
\end{equation}
where $N_{\nu}(t)$ is the number of neutrinos trapped in the system at the time $t$, and $\tau $ is a characteristic constant parameter. Carrying out the integration of equation (\ref{dNdt}), we get the neutrino number density at the time $t$:
\begin{equation}
n_{\nu} (t) = \frac{n_{\nu} (0)}{\rho (0)} \,  e^{- t/\tau} \rho (t),  
\end{equation}
where  $n_{\nu} (0)$ and $\rho (0)$ stand for the initial neutrino number density and the baryon density, respectively, and $\rho (t)$
is the baryon density at the time $t$.

The escape of neutrinos decreases the pressure of the core of the protoneutron star, which tends to contract due to the action of the inward gravitational force.  The pressure of the neutron star matter can be splitted into two terms 
\begin{equation}
P = P_0 + P_{\nu_e}, \label{pressure1}	
\end{equation}
where $P_0$ stands for the pressure of baryons, electrons and muons, and $P_{\nu}$ is the partial pressure of the remaining neutrinos that are still trapped in the core at the time $t$. The latter is given by  
\begin{equation}
P_{\nu}=\frac{1}{24\pi ^{2}}\mu_{\nu}^{4},
\end{equation}
where the neutrino chemical potential, $\mu_{\nu}$, reads 
\begin{equation}
\mu _{\nu }=\left( \frac{6}{\pi ^{2}} n_{\nu }\right) ^{1/3}.
\end{equation}

In writing equation (\ref{pressure1}), we are implicitly assuming that electrons and neutrinos are no longer in beta-equilibrium, which means that neutrinos decouple from the baryonic matter. Thus, the equation of state used to derive the pressure $P_0$ in equation  (\ref{pressure1}) can be computed separately or provided as input.

 The model parameter $\tau$ introduced in equation (\ref{dNdt}) governs the time scale of the escape of neutrinos, which is of the order of tens of seconds.

\section{Results and discussions}\label{sec:result}

 We assume that the spheroidal protoneutron star is initially composed of hadrons and leptons (including neutrinos), with the hadronic sector described in the framework of a relativistic mean field theory \citep{Walecka1974, Serot1986,glendenning00:book}. The initial values of the  semi-axes and the angular velocity are determined by a meta-stable spheroidal equilibrium configuration provided the mass $M$ and the initial angular momentum $L_0$ of the protoneutron star. To study the appearance of triaxial configurations, the initial values for the semi-axes $a_1$ and $a_2$ are slightly different in all variants of the calculations, in the eighth decimal place. Thus, we solve numerically the equations of motion (\ref{29})--(\ref{31}) for given initial equilibrium conditions, coupled with the evolution of the angular moment of the nascent neutron star in the equation (\ref{velocidadeangular}).
 
\begin{figure} [htbp] 
\centering
\includegraphics[width=.30\textheight]{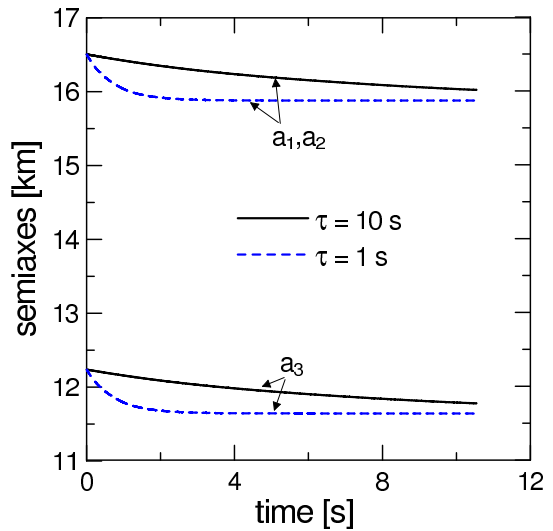}
\caption{(Color online) - Time evolution of the semi-axes of a spheroid of mass M = 1.6 M$_\odot$, and angular momentum $L_0 = 1.2 \times 10^{49}$ ${\rm g.cm^2.s^{-1}} $, for two different values of the neutrino escape parameter $\tau$, displayed in the figure.}
\label{Fig1}
\end{figure}

In order to illustrate the dynamical solution of the system, in Figure \ref{Fig1} it is shown the time evolution of the semi-axes of a  rotating $1.6$ $M_{\odot}$ protoneutron star. In this case, the initial angular momentum is $L_0 = 1.2  \times 10^{49} \, {\rm g.cm^2.s^{-1}} $. As we can see, the system maintains the spheroidal shape all the time, and thus no gravitational wave is generated.  

\begin{figure} [htbp] 
\centering
\includegraphics[width=.30\textheight]{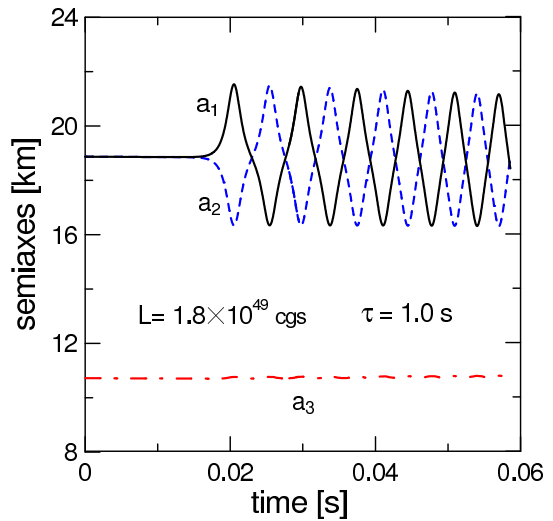}
\caption{(Color online) - First instants of the time evolution of the semi-axes showing the transition to the triaxial ellipsoid for the configuration of mass M = 1.6 M$_\odot$, $L_0 = 1.8 \times 10^{49}$ ${\rm g.cm^2.s^{-1}} $, and the neutrino escape parameter $\tau = 1.0$ s.}
\label{Fig2}
\end{figure}

The Figure \ref{Fig2} depicts the time evolution of the semi-axes of a rotating protoneutron star of mass $M=1.6$ $M_{\odot}$, but with higher initial angular momentum, namely $L_0 = 1.8  \times 10^{49} \, {\rm g.cm^2.s^{-1}} $. In this case due to the increasing in the initial rotation velocity we can see the axial symmetry breaking during first milliseconds of the evolution. The spheroid shape changes to a triaxial ellipsoid, as shown in the figure. At this point the system becomes a source of gravitational wave.

\begin{figure} [htbp] 
\centering
\includegraphics[width=.30\textheight]{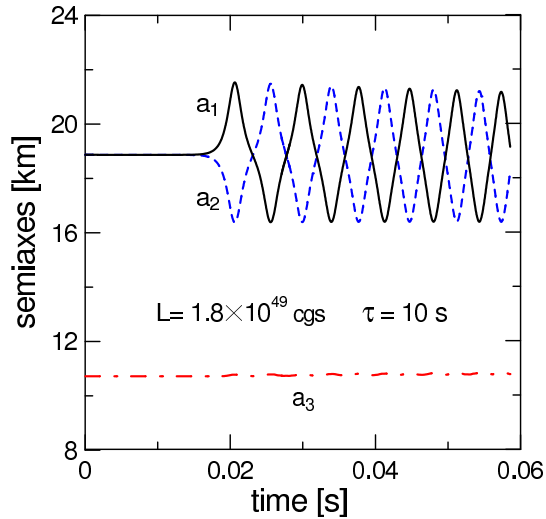}
\caption{(Color online) - Same as the previous Figure \ref{Fig2}, but for the neutrino escape parameter $\tau = 10.0$ s.}
\label{Fig3}
\end{figure}

Analogously, Figure \ref{Fig3} shows the three semi-axes as functions of time, but for the case of higher characteristic time scale of neutrino escape, $\tau = $ 10 s. The order of magnitude of the needed time for the occurrence of the shape transition is again smaller than the time scale of neutrino escape. For this and previous case the dynamical shape transition occurs with almost all neutrinos still trapped in the system, thus the instant of shape bifurcation and semi-axes evolutions are quite similar for both situations.


\begin{figure} [htbp] 
\centering
\includegraphics[width=.30\textheight]{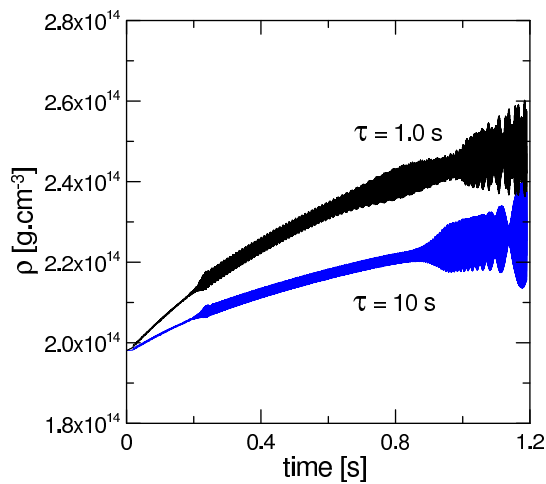}
\caption{(Color online) - Time evolution of the average density of the same configurations showed in Figures \ref{Fig2} and \ref{Fig3}. }
\label{Fig4}
\end{figure}

In the Figure \ref{Fig4} we represent the time evolution of the average density of the configuration depicted in the Figures \ref{Fig2} and \ref{Fig3}, for the two values of the parameter $\tau$. The average bulk density reaches values greater than the normal nuclear density, and increases in time due mainly to the average core contraction caused by the loss of rotational energy carried by the gravitational wave.

\begin{figure} [htbp] 
\centering
\includegraphics[width=.30\textheight]{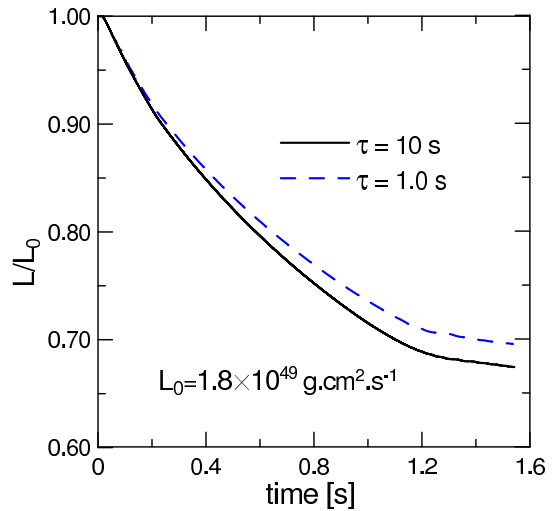}
\caption{(Color online) - Time evolution of the angular momentum of the same configurations showed in figures 2 and 3. }
\label{Fig5}
\end{figure}

Figure \ref{Fig5} shows the lost of angular momentum as a function of time for the evolutions shown in Figures \ref{Fig2} and \ref{Fig3}.  The rate of angular momentum loss due to the gravitational radiation is governed by Equation (\ref{63}). 

\begin{figure} [htbp] 
\centering
\includegraphics[width=.30\textheight]{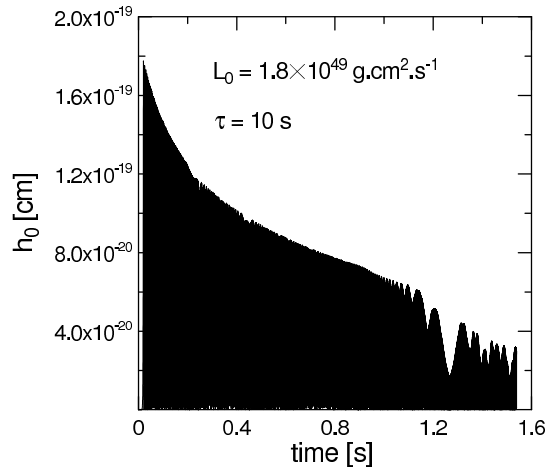}
\caption{Amplitude of the gravitational wave emitted by a source distant $10$ kpc from the Earth, corresponding to the dynamical evolution presented in the Figure \ref{Fig3}.}
\label{Fig6}
\end{figure}

At this point, it is important to consider some observational aspects of the  gravitational wave generated in the aftermath of the dynamical shape transition. For this purpose, in Figure \ref{Fig6} we show the behavior of the amplitude of the emitted gravitational radiation considering the source distant $10$ kpc from the Earth, relative to the time evolutions depicted in Figure \ref{Fig3}. The obtained amplitudes are within the range of values detectable with current techniques for pulses with periods of short duration, of the order of $10^{-22}$--$10^{-21}$ cm \citep{Ohanian}. 

\begin{figure} [htbp] 
\centering
\includegraphics[width=.30\textheight]{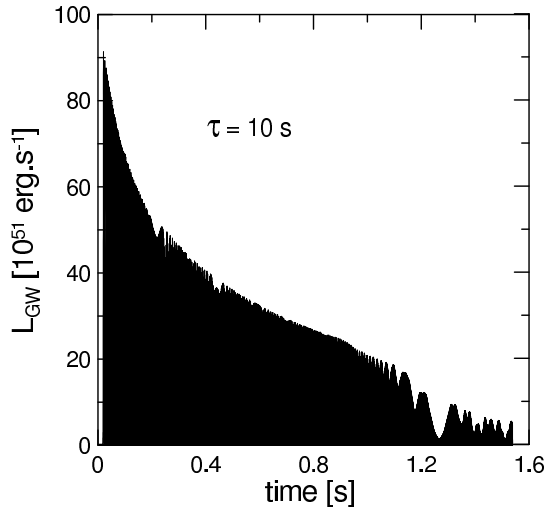}
\caption{Luminosity of the gravitational waves emitted by the configuration showed in the Figure \ref{Fig3}. }
\label{Fig7}
\end{figure}

\begin{figure} [htbp] 
\centering
\includegraphics[width=.30\textheight]{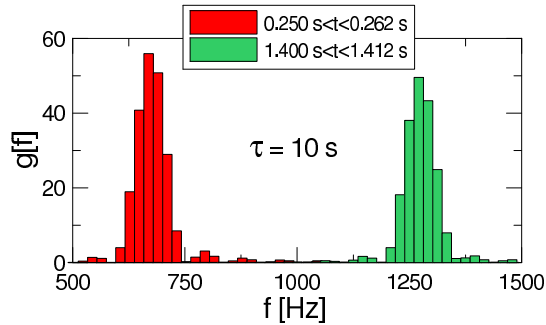}
\caption{(Color online) - Power spectrum for two different time intervals computed for the luminosity of the gravitational waves showed in Figure \ref{Fig7}.  }
\label{Fig8}
\end{figure}

In Figure \ref{Fig7} it is depicted the gravitational luminosity as a function of time also for the model calculation in Figure \ref{Fig3}. The corresponding power spectrum of the emitted gravitational wave (using two different time intervals to the frequency decomposition) is shown in Figure \ref{Fig8}. The spectrum is very sharp and we can estimate that the characteristic frequencies range from about 670 Hz to 1270 Hz in this case.


\begin{figure} [htbp] 
\centering
\includegraphics[width=.30\textheight]{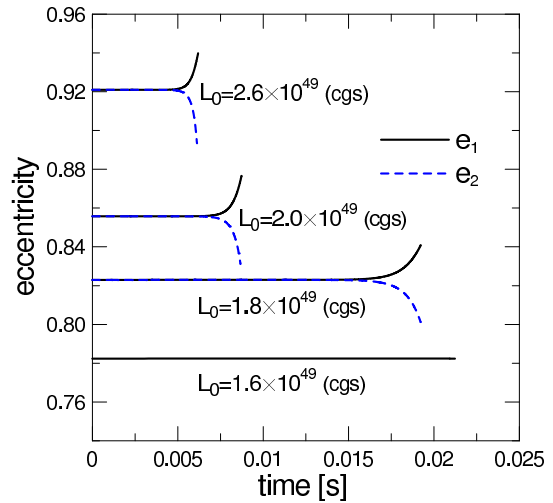}
\caption{(Color online) - Time evolution of the equatorial eccentricities $e_1$ (solid lines) and $e_2$ (dashed lines), for some values of the initial angular momentum. The axial symmetry breaking occurs from a critical value of the initial angular momentum (close to $L_0 = 1.8  \times 10^{49} \, {\rm g.cm^2.s^{-1}} $, for the 1.6 $M_\odot$ spheroid.). }
\label{Fig9}
\end{figure}

\begin{figure} [htbp] 
\centering
\includegraphics[width=.30\textheight]{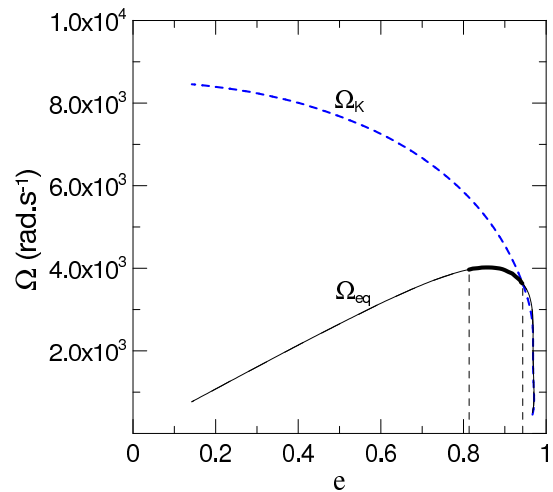}
\caption{(Color online) - Relation between the initial angular frequency and eccentricity of the spheroids (solid line). The upper bound of the rotational frequency is determined by the Kepler frequency (dashed line). The small thick line indicates the possible values of angular frequency (angular momentum) which allow the shape transitionn.}
\label{Fig10}
\end{figure}

We remark from comparison of Figures \ref{Fig1}-\ref{Fig3} that the relevant aspect to the occurrence of the dynamical shape transition in the early stage of the time evolution of the pulsar core is the initial value of the angular momentum of the nascent pulsar. In Figure \ref{Fig9} we illustrate this aspect showing the time evolution of the equator eccentricities $e_1$ and $e_2$, defined by $e_i =\sqrt{1-a_i^2/a_3^2}$ $(i=1,2)$, for four different values of the initial angular momentum, and for the mass 1.6 $M_\odot$. Notice that the bifurcation to the triaxial configuration is reached at some critical value of the initial equatorial eccentricity, around  $e=0.82$ (which corresponds to a critical value of initial angular momentum close to $L_0 = 1.8  \times 10^{49} \, {\rm g.cm^2.s^{-1}} $). This means that when the rotational symmetry of the spheroid is dynamically broken, the system releases more degree of freedom in its evolutionary path. The asymmetry of the triaxial ellipsoid creates the necessary mass quadrupole to generate the gravitational wave.

The small thick part in the Figure \ref{Fig10} represents the allowed values of initial angular frequency and eccentricity of the spheroid, capable to generate the axial symmetry breaking and so the generation of gravitational wave. This set of values should depend on the used value of the pulsar core mass.

\section{Conclusions}\label{sec:conclu}

In summary, in this work we have presented a model of gravitational wave generated by the core of rotating protonneutron stars. The equilibrium of the initial configuration is disrupted by the emission of neutrinos formerly trapped during the gravitational collapse, which suffices to trigger the dynamical evolution of the system. Starting from spheroidal equilibrium configurations, we show that the axial symmetry is dynamically broken for high values of angular momentum, and that the system evolves for a fast rotating triaxial ellipsoid, emitting gravitational radiation.

 This happens only for initial configuration of the spheroidal core in a window of angular frequency for the given core mass. With the dynamically broken spheroidal symmetry of the pulsar its become a source of gravitational wave. The obtained configurations are uniformly rotating objects with period, mean radius, density and chemical composition typical of observed pulsars. The luminosity, the frequency and the amplitude of the radiated gravitational wave are estimated, showing that they are in the range of the detectable values expected for next generation of gravitational wave detectors.

\acknowledgments S. B. Duarte thanks CNPq for financial support.

\end{document}